\documentclass[preprint,preprintnumbers,amsmath,amssymb,superscriptaddress]{revtex4}

\usepackage{graphicx}
\usepackage{dcolumn}
\usepackage{bm}
\usepackage{subfigure}
\usepackage{placeins}
\usepackage{multirow}
\usepackage{amsmath}
\usepackage{slashbox}
\usepackage{multirow}

\begin{document}


\title{The extreme vulnerability of interdependent spatially embedded networks}

\author{Amir Bashan}
\affiliation{Minerva Center and Department of Physics, Bar Ilan University,
Ramat Gan, Israel.}

\author{Yehiel Berezin}
\affiliation{Minerva Center and Department of Physics, Bar Ilan University,
Ramat Gan, Israel.}

\author{Sergey V. Buldyrev}
\affiliation{Department of Physics, Yeshiva University, New York, New York 10033, USA.}

\author{Shlomo Havlin}%
\affiliation{Minerva Center and Department of Physics, Bar Ilan University, Ramat Gan, Israel.}

\date{\today}

\maketitle

\textbf{
Recent studies show that in interdependent networks a very small failure in one network may lead to catastrophic
consequences. 
 Above a critical fraction of interdependent nodes, even a single node failure can invoke cascading failures that may abruptly
fragment the system, while below this
"critical dependency" (CD) a failure of few nodes leads only to small damage to the system. 
So far, the research has been focused on interdependent random networks without space limitations. 
However, many real systems, such as power grids and the Internet, are not random but are spatially embedded.
Here we analytically and numerically analyze the stability of systems consisting of
interdependent spatially embedded networks modeled as lattice networks.
Surprisingly, we find that in lattice systems, in contrast to non-embedded systems, there is no CD and \textit{any} small fraction of interdependent nodes leads to an abrupt collapse.
We show that this extreme vulnerability of very weakly coupled lattices is a consequence of the critical exponent
 describing the percolation transition of a single lattice.
Our results are important for understanding the vulnerabilities and for designing robust interdependent spatial embedded networks.
}

Complex systems, usually represented as complex networks, are rarely isolated but usually
 interdependent and interact with other systems \cite{Rosato,Peerenboom,Rinaldi}.
Recently it was shown that a coupled networks system is considerably more vulnerable than its isolated
component networks \cite{Buldyrev2010,Parshani2010prl,Vespignani2010,Parshani2010epl,Shao2011, Leicht2011, D'Souza2011, Huang2011, Gao2011, Sergey2011pre, Hu2011, Xu2011, Hao2011, Gu2011, BashanNatComm, Parshani2011, Bashan2011,Gao2010arxiv, Bashan2011pre}.
A failure of nodes in one network leads to a failure of dependent nodes in other networks,
 which in turn may cause further damage to the first network, leading to cascading failures and
 catastrophic consequences.
It was shown that the strength of the coupling between the networks, represented by the fraction $q$ of
interdependent nodes, determines the way the system collapses \cite{Gao2011,Parshani2010prl,GaoNaturePhysics}.
For strong coupling, that is for high fraction of interdependent nodes, an initial  damage can lead to
cascading failures that yield an abrupt collapse of the system,
 in a form of a first-order phase transition.  Reducing the coupling strength
 below a critical value, $q_c$, leads to a change from an abrupt collapse to continuous decrease of
  the size of the network, in a form of a second-order phase transition.
This new paradigm is in marked contrast to the common knowledge represented by a single network behavior.
In a single network a failure of few nodes can make only a small damage to the
 network, while a system of interdependent networks might be functioning  well with most nodes connected, but a further failure of even a single node may lead to
 a complete collapse of the entire system.
Thus, the existence of an abrupt collapse phenomena in interdependent networks makes such systems extremely risky. Thus, understanding this phenomena is critical for evaluating the systems' vulnerability and for designing robust infrastructures.

Current models focus on interdependent networks where space restrictions are not considered.
Indeed, in some complex systems the spatial location of the nodes is not relevant or
not even defined, such as in proteins interaction networks \cite{Milo2002,Alon2003,Khanin} and the World Wide Web \cite{Cohen2000,Dorogovtsev}.
However, in many real-world systems, such as power grid networks and computer networks,
nodes and links are located in Euclidian two-dimensional space. The dimension and universality class
of embedded networks, whose links have a characteristic length, is the same for all embedded networks \cite{HavlinBook}.
For example, in power grid networks the links have a characteristic length since their lengths follow an exponential distribution
\cite{Li2011dimension}. Thus, to obtain the main features of the system under failures, we
model such a spatial embedding network as a two dimensional lattice.

Here, we study the case of interdependent lattice networks where a fraction $q$ of nodes in each lattice
randomly depends on nodes  in other networks.
We find that in the case of coupled lattices $q_c=0$ and any coupling $q>0$ leads to a first-order transition.
We show that the origin for this extreme vulnerability of coupled lattices compared to random networks $(q_c>0)$
lies in the critical behavior of percolation in a single lattice, which is characterized by the
 critical exponent $\beta <1$, in contrast to random networks where $\beta=1$ \cite{Stauffer}.
 Our theoretical and numerical approaches predict that a real-world system of interdependent spatially embedded networks which are characterized by $\beta<1$ will, for any $q>0$, abruptly disintegrate.
Since for percolation of lattice networks it is known that for any dimension $d<6$, $\beta<1$ \cite{HavlinBook}, we expect that also interdependent systems embedded in $d=3$ (or any $d<6$) will collapse abruptly for any finite $q$.

In addition of studying a pair of interdependent lattices we also analyze the stability of a more general case of
a starlike network-of-lattices, where each of the $n-1$ peripheral lattices is coupled to the root lattice with the
same coupling strength $q$.
We find that the root lattice abruptly collapses (first-order transition) for any $q$ and any $n>1$, however, the percolation behavior of the peripheral
 lattices is somewhat counterintuitive:
for small $n$ the peripheral lattices also abruptly collapse together with the root lattice,
while for large $n$ the peripheral lattices may remain functional even after the root collapses.


\begin{figure}[ht]
 \begin{center}
      \subfigure{\includegraphics[width = 0.350\textwidth]{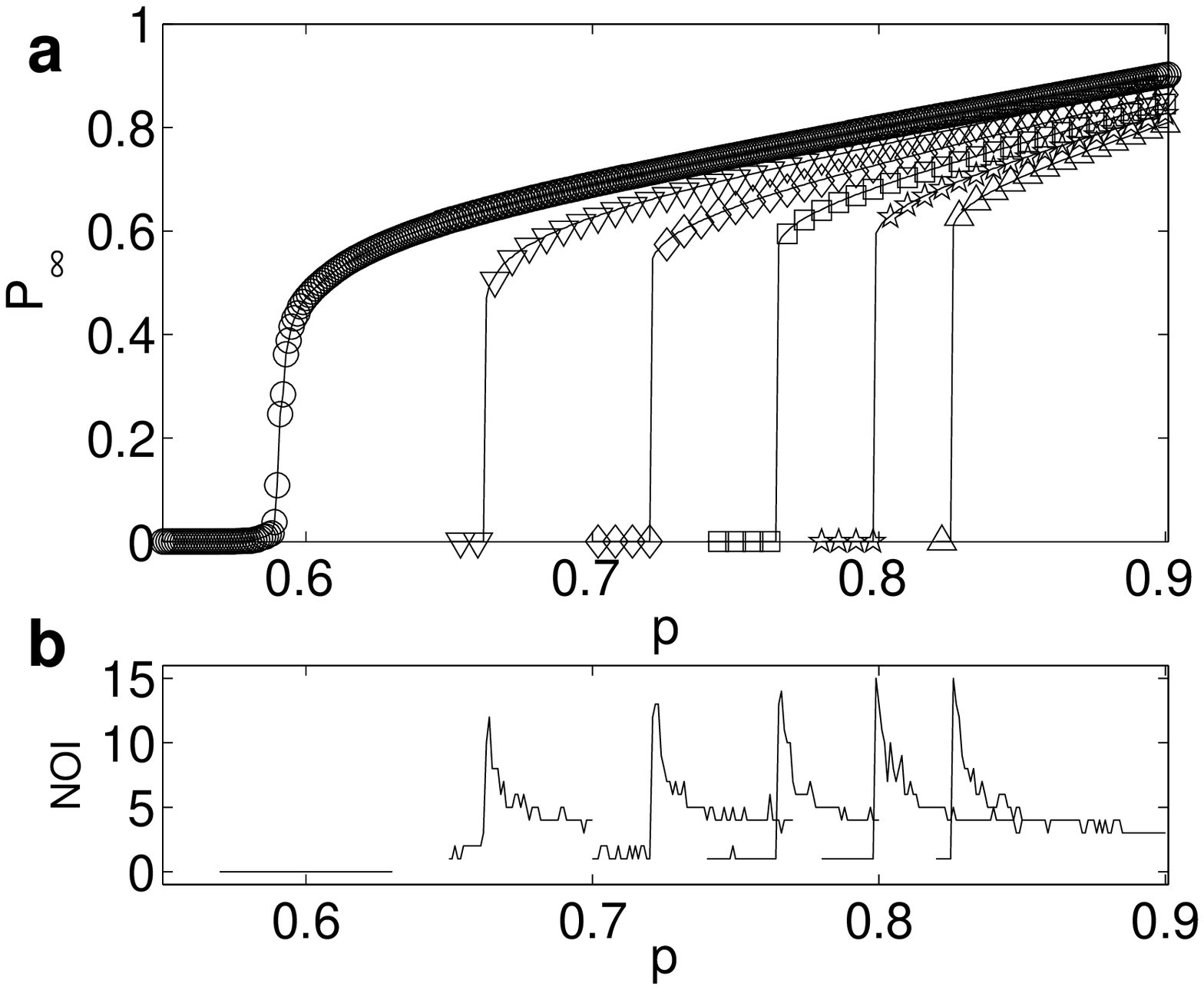}}
      \subfigure{\includegraphics[width = 0.350\textwidth]{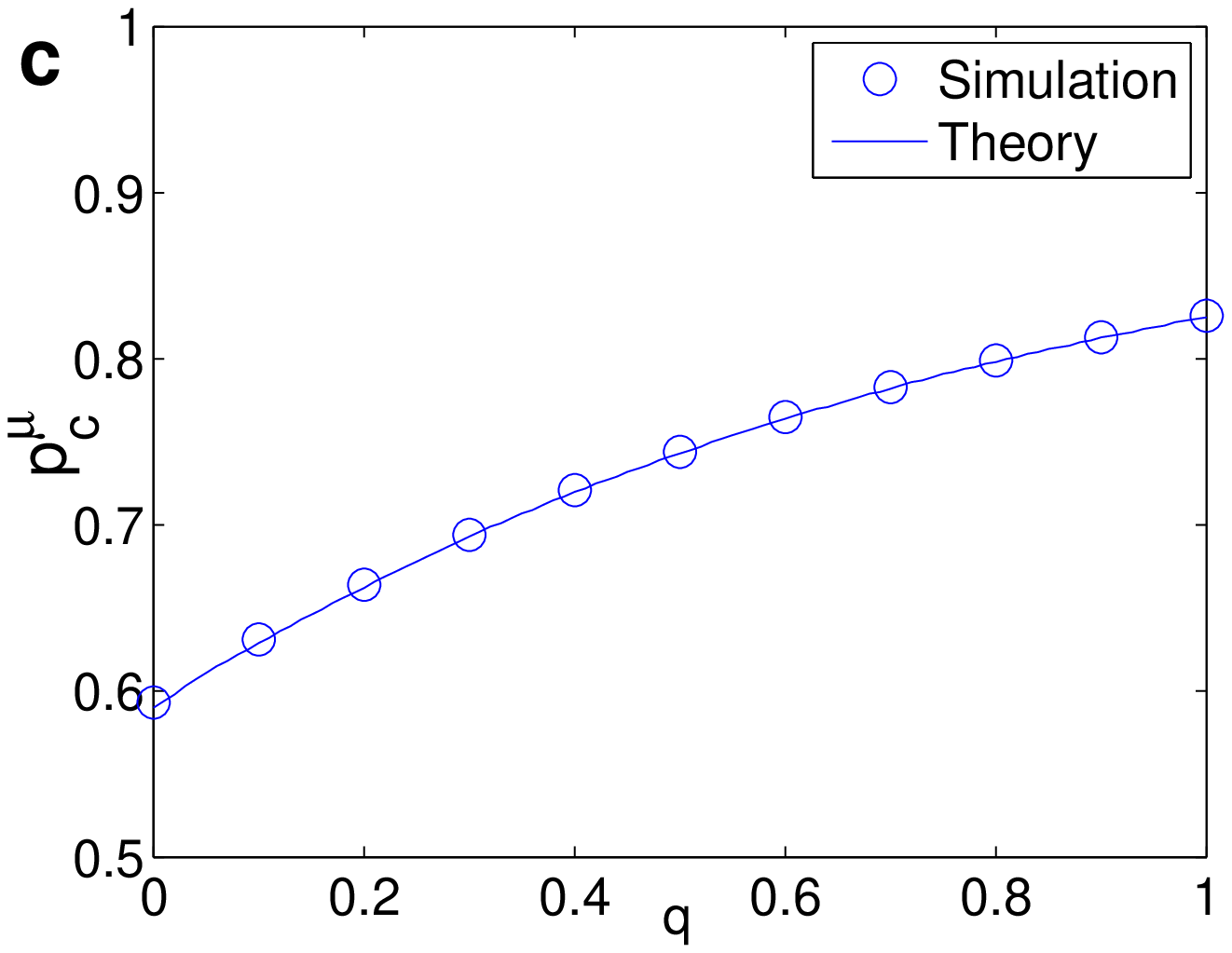}}
  \end{center}
    \caption{\textbf{Abrupt collapse of interdependent lattices.} \textbf{a,} The size of the giant component
    $P_\infty$ of two interdependent lattice networks, each of size $4000\times 4000$, at steady state after random failure of $1-p$ of their nodes.
     The solid lines are the solutions of Eq. (\ref{eq_symmetric}) and the symbols represent simulation results.
     In the case of $q=0$ (circles), no coupling between the networks, the transition is the conventional second-order percolation,
     while for any $q>0$ (triangle downs, $q=0.2$; diamonds, $q=0.4$; squares, $q=0.6$; stars, $q=0.8$; triangles up, $q=1$) the collapse is abrupt in the form of first-order transition.
     \textbf{b,} The number of iterations (NOI) vs. $p$, triggered by removing at each step a fraction $\Delta p=10^{-3}$ of the nodes.
      In the case of first-order transition  (abrupt collapse due to cascading of failures), the number of
     cascading failures, NOI, diverges for $p\rightarrow p_c^\mu$, providing a useful approach for detecting accurately the critical
     threshold $p_c^\mu$ in simulations \cite{Parshani2011}.
     \textbf{c,} The solid line is the solution of Eqs. (\ref{eq_symmetric}) and (\ref{eq_der_symmetric}) for $p_c^\mu$ versus $q$ and the symbols represent results from simulations.
    Note that for the case of no coupling between the networks,
 $q=0$, we obtain the known result of a single lattice $p_c^\mu=0.593$ and for the case of complete coupling $q=1$ we obtain $p_c^\mu=0.826$,
 in agreement with \cite{WeiPRL}.
    \label{p_inf_vs_p}}
\end{figure}

Our analytical considerations consist of four stages.
First, we analyze the conditions for having an abrupt transition (first-order percolation transition) in a system of any coupled networks system.
Second, we find the critical coupling strength $q_c$ below which the transition becomes continuous (second-order).
Third, we show that for the case of two coupled lattices $q_c=0$, thus the transition is of first-order for any $q>0$.
Finally, we generalize our results to the case of network-of-lattices.

\section{Symmetric interdependent networks}
\label{SymmNofeedback}
\begin{figure}[ht]
   \begin{center}
      \subfigure{\includegraphics[width = 0.20\textwidth]{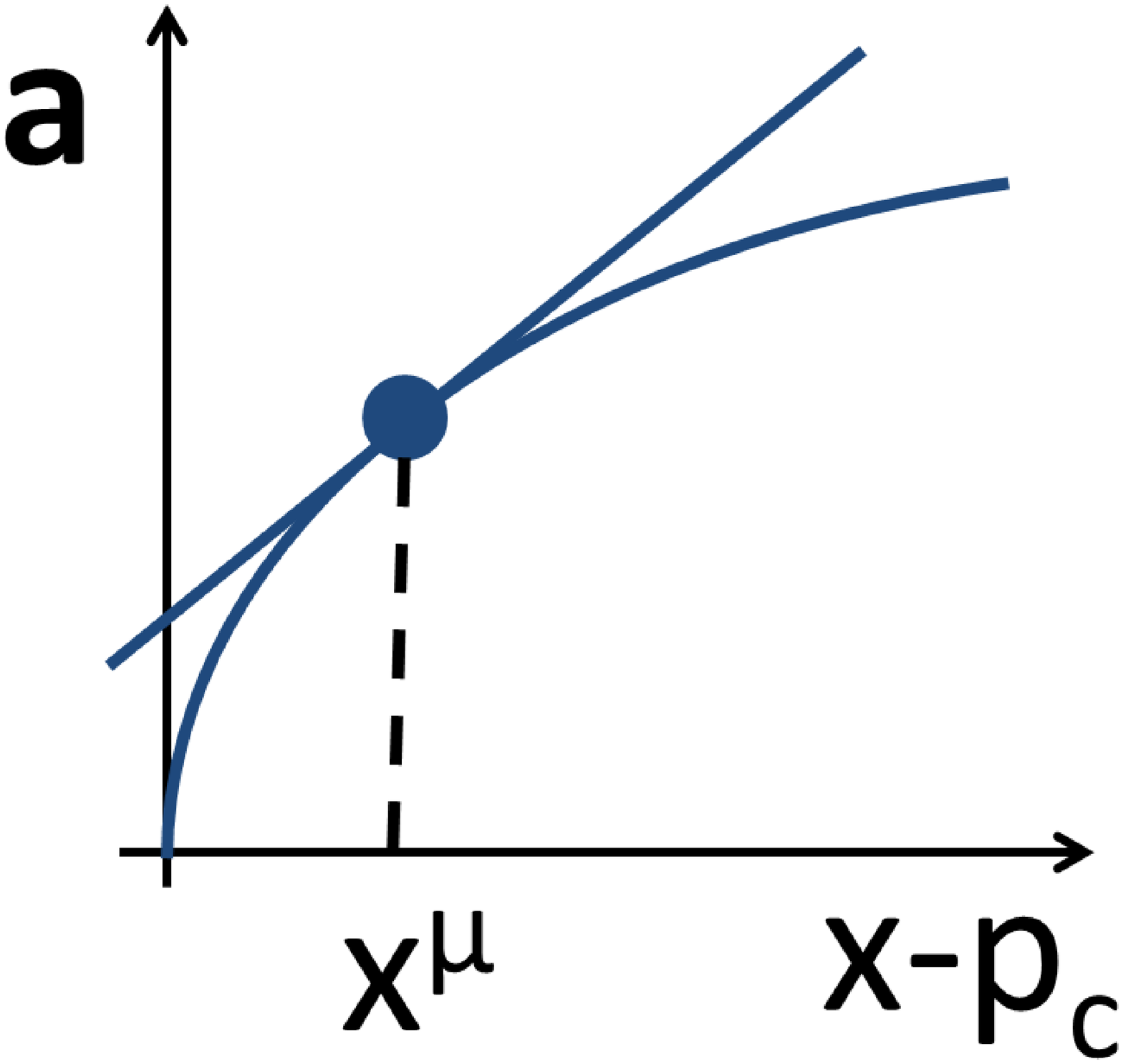}}
      \subfigure{\includegraphics[width = 0.20\textwidth]{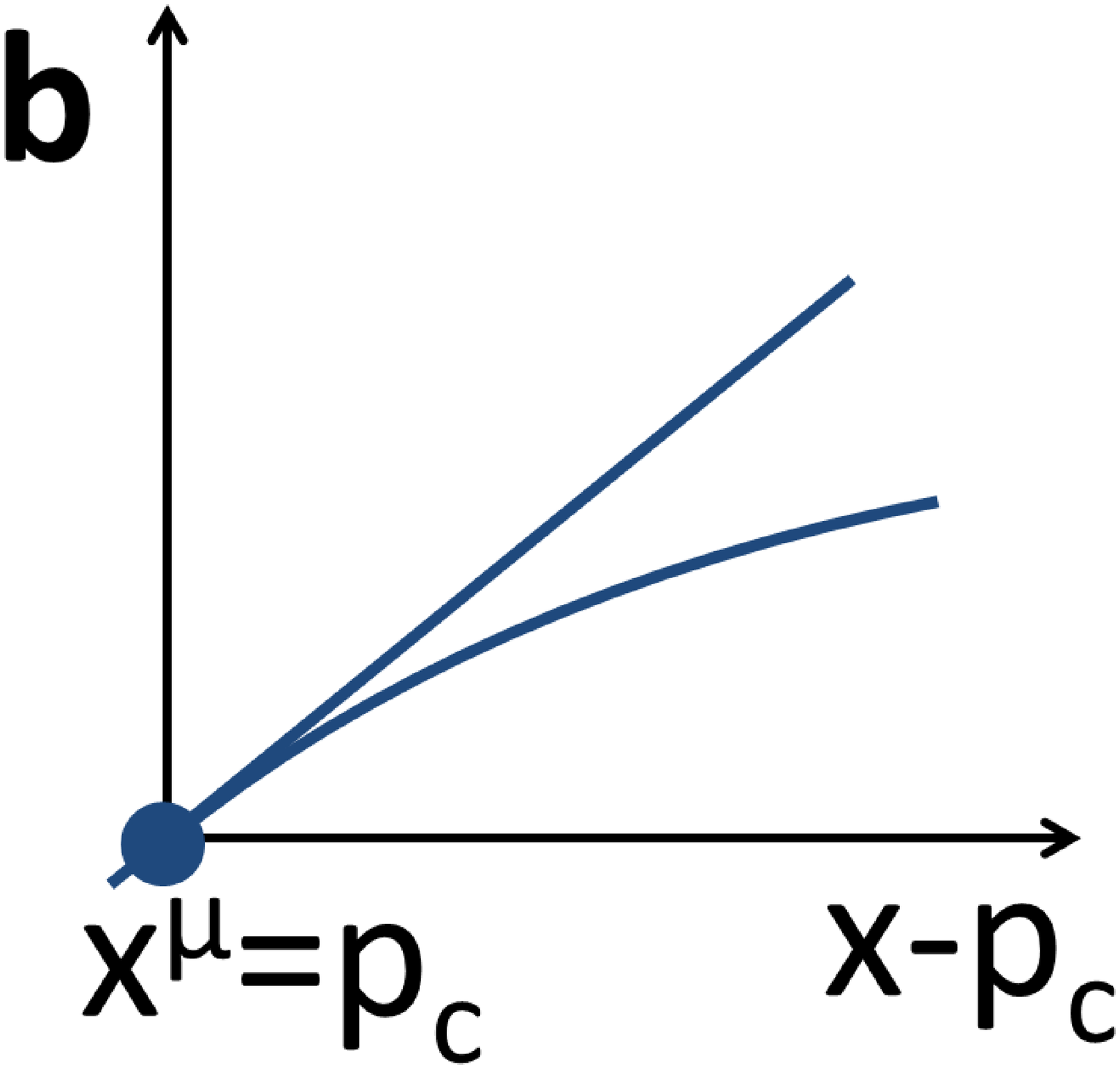}}
      \\
      \subfigure{\includegraphics[width = 0.20\textwidth]{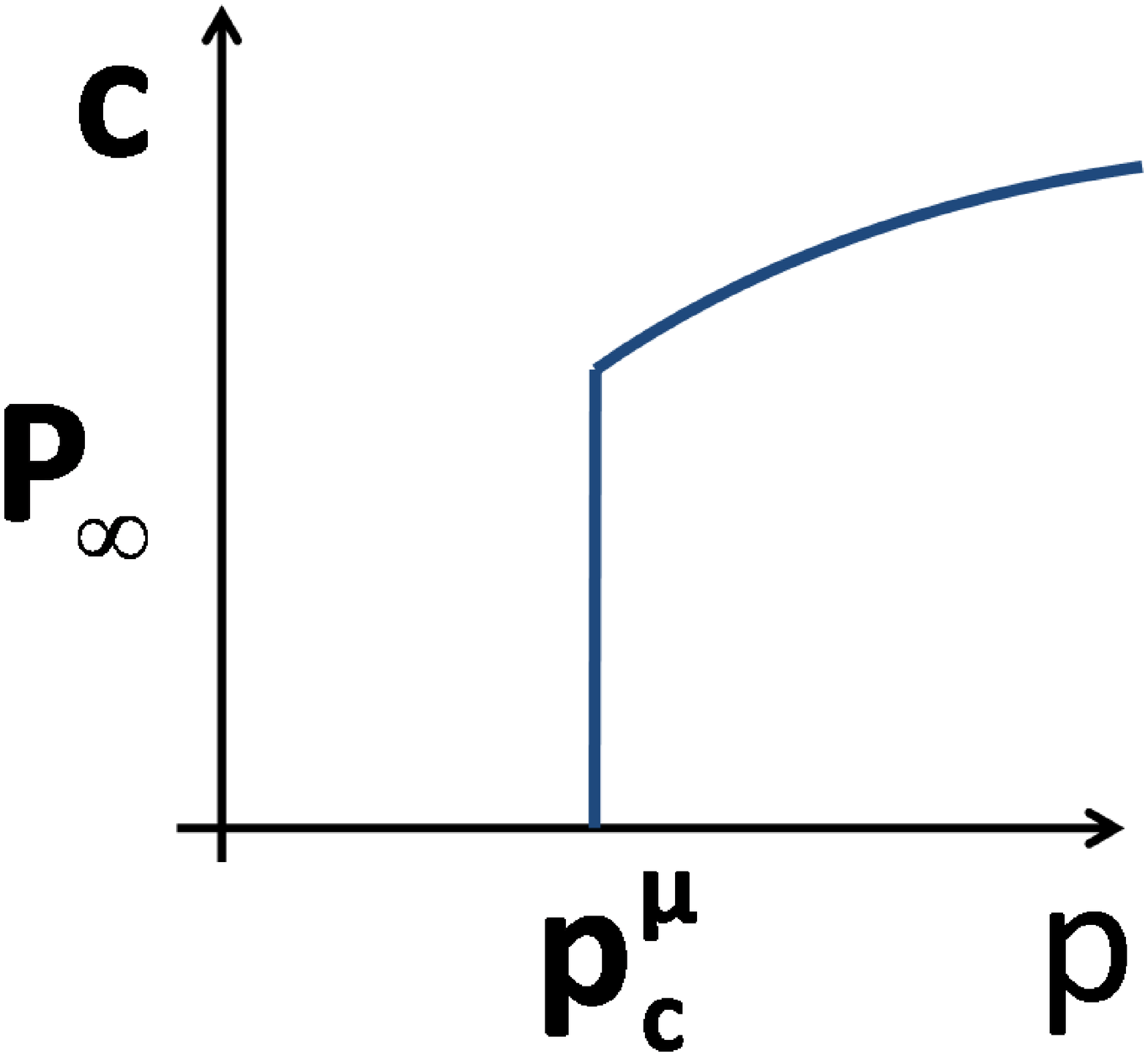}}
      \subfigure{\includegraphics[width = 0.20\textwidth]{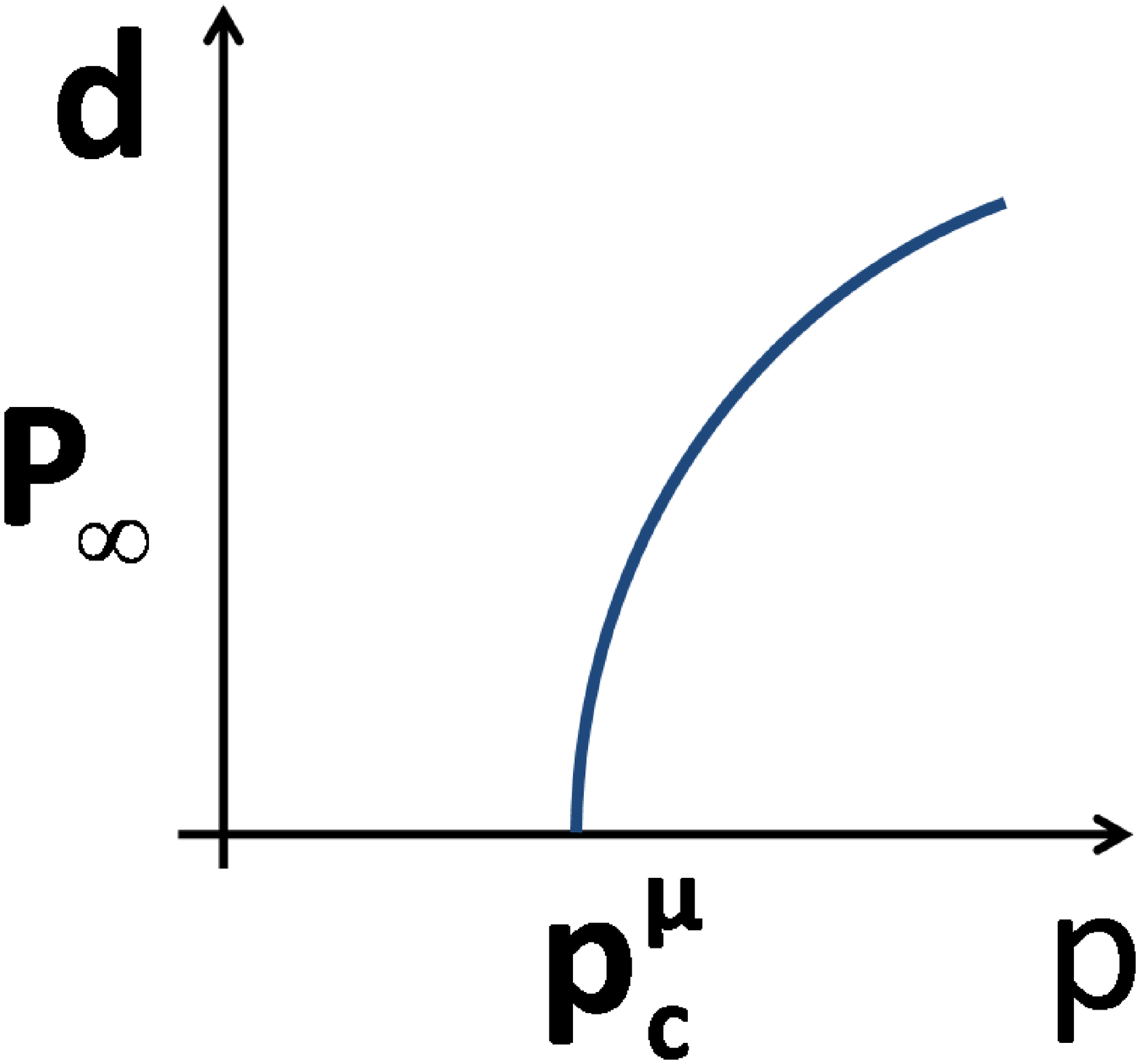}}
   \end{center}
   \caption{Schematic solution of Eq. (\ref{eq_symmetric}) for $p=p^\mu_c$ in the case of two coupled (a)
lattices and (b) ER networks. The mixed ER-Lattice case is studied by Landau et. al. \cite{Landau}.
 The left-hand-side (straight line) and right-hand-side (curve)
 of Eq. (\ref{eq_symmetric}) are plotted versus $x$, where the solution at criticality ($x=x^\mu$)
 is marked as solid circle. For the case of (a) two coupled lattices, since $g'(p_c)\rightarrow \infty$ the value
of $x^\mu$ is always larger than $p_c$, thus, the transition is of first-order for any non-zero $q$ value (c).
However, for the case of (b) such as for two ER networks, $g'(x)$ is finite for any $x$. Thus, for $q<q_c$
there is no tangential touching point between the straight line and the curve and the solution $x$ continuously
approaching $p_c$ as $p$ decreases so the transition is of second-order (d).}
\label{schematic}
\end{figure}

We analyze here, for simplicity and without loss of generality, the symmetric case where the two networks, A and B, have the same degree
distribution. A fraction $q$ of the nodes of each network are dependent on nodes randomly selected in the other network
under the assumption of no-feedback condition, meaning that if node $A_i$ depends on node $B_j$ , and $B_j$ depends on $A_k$ , then
$k=i$ \cite{Buldyrev2010}.
The analysis of the non-symmetric case and the case with feedback which yield similar results are discussed in the sections \ref{NonSymmSect}
and \ref{feedback} respectively.
In the symmetric case a fraction $1-p$ of nodes of both
networks is initially randomly removed. As a result, a certain fraction of nodes become disconnected from each network,
while a fraction $g(p)$ of the nodes in each network remains in the giant component.
Each node that has been removed or disconnected
from the giant component causes its dependent node in the other network to also fail. This leads to further
disconnections in the other network and to cascading failures. The size of the networks' giant components at steady state
 is given by $P_\infty=xg(x)$, where $x$ is the solution of the self consistent equation \cite{Gao2011}
\begin{equation}\label{eq_symmetric}
    x-p(1-q)=p^2qg(x).
\end{equation}
The function $g(x)$ represents the probability of a node to be connected to the giant component after random removal of a
fraction $1-x$ of the nodes and can be obtained either analytically or numerically from the percolation behavior of a \textit{single} network.
In general, $g(x)$ has a critical value at $x=p_c$ such that $g(x\leq p_c)=0$ while $g(x>p_c)>0$ and monotonically increases with $x$.
For any given value of $p$, a graphical solution of Eq. (\ref{eq_symmetric}) is given as the intersection
of the straight line $y=x-p(1-q)$ and the curve $y=p^2qg(x)$. The value of $p$ where the line and the curve tangentially intersect,
$p \equiv p^\mu_c$, corresponds to a discontinuity in the solution of $x$ (see Fig. \ref{schematic}a). This leads  also to a discontinuity in the solution of $P_\infty(p)$
which abruptly jumps to zero as $p$ slightly decreases.
At any $p<p^\mu_c$, $x$ is smaller than the critical value $p_c$ (of a single network) and therefore $P_\infty(p<p^\mu_c)=0$.

In Fig.~\ref{p_inf_vs_p}(a) we show $P_\infty(p)$ from theory, Eq. (\ref{eq_symmetric}), and simulations as a function of $p$
for various values of $q$. For $q=0$ the lattices are not coupled and the percolation transition is the conventional continuous second-order, while for $q>0$ the transition is discontinuous first-order. As seen, the theory and simulations are in excellent  agreement. In order to find in simulations $p_c^\mu$ accurately
we evaluate the number of cascade iterations, (NOI) \cite{Parshani2011}, for each $p$, which has a sharp peak (diverges for infinite systems) at $p_c^\mu$ as seen in Fig. \ref{p_inf_vs_p}(b).

The condition for the abrupt jump is that additional to (\ref{eq_symmetric}), the derivatives of both sides of Eq. (\ref{eq_symmetric})
 with respect to $x$ are equal,
 \begin{equation}\label{eq_der_symmetric}
    1=p^2qg'(x).
 \end{equation}
Solving Eq. (\ref{eq_symmetric}) together with Eq. (\ref{eq_der_symmetric}) for a given $q$ provides the abrupt jump transition
point $p^\mu_c$ and $x^\mu$, which yields the size of the giant component just before collapsing, $P_\infty^\mu\equiv x^\mu g(x^\mu)$.

The size of the giant component at criticality $P^\mu_\infty$ depends on the coupling strength $q$ such that
reducing $q$ leads to smaller value of $x^\mu$ and thus smaller $P_\infty^\mu$.
As long as $x^\mu>p_c$, $P_\infty^\mu>0$ and $P_\infty(p)$ discontinuously jumps to zero at $p$ just below $p^\mu_c$.
However, for a certain $q$, $x^\mu \rightarrow p_c$ and the size of the jump becomes zero since $P_\infty^\mu \rightarrow 0$
(see Fig. \ref{schematic}b,d).
In this case the percolation transition becomes continuous.
Therefore, the critical dependency $q_c$ below which the discontinuous transition becomes continuous,
must satisfies both Eq. (\ref{eq_symmetric}) and Eq.(\ref{eq_der_symmetric}) for $x\rightarrow p_c$, and
is given by
\begin{eqnarray}\label{eq_symm_qc}
  p(1-q_c)&=& p_c\\
  \nonumber p^2q_c g'(p_c) &=& 1 .
\end{eqnarray}
The set of equations (\ref{eq_symm_qc}) is a general condition for finding the transition point, $q_c$, from discontinuous
 to continuous percolation transition in a symmetric case of any two networks. Interestingly, the transition point, $q_c$, depends only on
the single network behavior near criticality, $g(x=p_c)$, and not on the entire shape of $g(x)$.
As seen in the following, in the case of random networks
$g'(p_c)$ is finite yielding a finite solution for $q_c$. However, for the case of lattice networks
the derivative of $g(x)$ diverges at the critical point, $g'(p_c)= \infty$, yielding $q_c=0$.
Therefore, from Eq. (\ref{eq_symm_qc}) follows that any coupling between lattices leads to an abrupt first order transition.

The behavior of the percolation order parameter of a single network near the critical point is defined by the critical exponent
$\beta$, where $g(x)_{x \rightarrow p_c}=A(x-p_c)^\beta$. The divergence of $g'(x)$ of 2d single lattice when
$x \rightarrow p_c$ is since $\beta=5/36<1$ \cite{MPM1979,Nienhuis1982,HavlinBook}.
In contrast, for ER networks $\beta=1$ which yields finite value of $g'(p_c)$.
Thus, $q_c$ can be explicitly calculated for coupled ER networks with average degree $\langle k \rangle$ from Eqs. (\ref{eq_symm_qc}) based solely on the behavior at criticality,
$p_c=1/\langle k \rangle$ and $A=2\langle k \rangle$, yielding
$q_c=1-\frac{1}{\langle k \rangle}(\sqrt{2\langle k \rangle +1}-1)$.%

\begin{figure}[ht]
    \includegraphics[width = 0.5\textwidth]{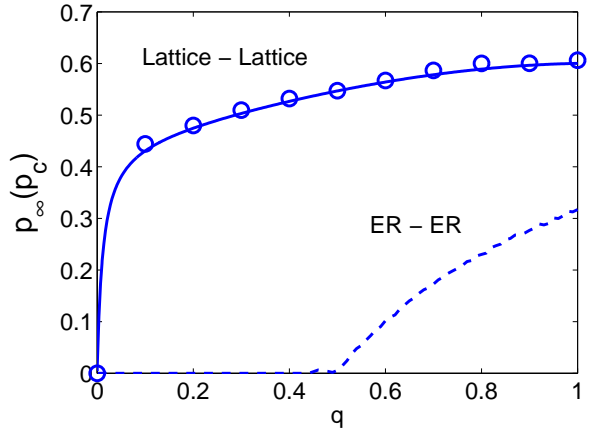} 	
    \caption{\textbf{The size of the collapse in coupled lattices compared to coupled random networks.}
    Comparison of the size of the giant component at criticality $P_\infty(p^\mu_c)$ for coupled lattices (circles) and
    coupled random (ER) networks with average degree $\langle k\rangle =4$ (squares) versus the coupling strength $q$.
    While for random networks with $q<q_c=0.5$ the size of the networks at criticality is zero, in coupled lattice network
    the networks abruptly collapse for any finite $q>0$.
    Note also the huge differences in the network sizes at the collapse transition. The coupled lattices collapse at a significantly larger giant components compared to the ER case.
    The solid line represent the theory for coupled lattices given by Eqs. (\ref{eq_symmetric})
    and (\ref{eq_der_symmetric}), and the circles are from simulations. The dashed line represents the theory for ER coupled networks derived in Parshani et. al. \cite{Parshani2010prl}.
    \label{Critical_Pinf_vs_q}}
\end{figure}

Figure \ref{schematic} demonstrates schematically the possible solutions of Eq. (\ref{eq_symmetric}) at criticality, $p=p_c^\mu$,
where the value of $x^\mu$ yields the size of the system just before the abrupt collapse, $P^\mu_\infty=x^\mu g(x^\mu)$.
While for (b) coupled ER networks and $q=q_c$ the solution is $x^\mu=p_c$ and $P^\mu_\infty=0$, meaning that
the system continuously disintegrates, for (a) coupled lattices, $x^\mu>p_c$ and $P^\mu_\infty>0$ for any $q$, thus,
the system of coupled lattices always undergoes a first-order transition.
Abrupt collapses are extremely risky since most of the system's elements just before the collapse may be functioning well
and no sign of warning appear. This is in mark contrast to single network behavior where a failure of small fraction of nodes can lead only to a small damage to the network.
Next, we study the size of the
giant component of coupled lattices at criticality $P_\infty(p_c^{\mu})$ for different coupling strengthes $q$, using
Eqs. (\ref{eq_symmetric}) and (\ref{eq_der_symmetric}) as well as numerical simulations. As shown in Fig. \ref{Critical_Pinf_vs_q},
in the case of a coupled lattices system for any $q>0$, $P_\infty(p_c^{\mu})>0$, that is the size of the collapse, is larger than zero for any coupling between the lattices. We also compare, in Fig. \ref{Critical_Pinf_vs_q}, the size of the collapse in a system of coupled lattices and
 in a system of coupled random ER networks with average degree $\langle k\rangle=4$, same as the degree of the lattice.
 Figure \ref{Critical_Pinf_vs_q} demonstrates the significantly increased vulnerability of the coupled lattices system compared with the random networks system. For example in the coupled random networks, for $q=q_c=0.5$
 the size of the system at criticality is zero, while in coupled lattices the collapse occurs when the giant component is about $1/2$ of the original network.

Following a similar approach, we analyze in section \ref{NonSymmSect} the non-symmetric case of two interdependent networks,
where a fraction $q_{ij}$ of the nodes of network $j$ depend on nodes of network $i$.
In section \ref{NonSymmSect} we show analytically that the first lattice that breaks down undergoes always a first-order percolation transition,
for any finite values of $q_{12}$ and $q_{21}$.
Yet, the percolation behavior of the second lattice, does depend on $q_{12}$ and falls into one of two scenarios:
 the second lattice either (i) abruptly collapses together with the first lattice at $p^{\mu}_{c2}=p^\mu_{c1}$
or (ii) undergoes a second-order transition at $p^{\mu}_{c2}< p^\mu_{c1}$, but still there is an abrupt fall at $p_{c1}^\mu$.

When the first lattice completely collapses at $p=p^\mu_{c1}$ the cumulative failures of the second lattice can
be evaluated by an effective occupation probability
\begin{equation}\label{eq_p_eff}
    p_{eff}=p^\mu_{c1}(1-q_{12}).
\end{equation}
When $q_{12}\geq 1-\frac{p_c}{p^\mu_{c1}}$ then $p_{eff}<p_c$ and the two lattices will abruptly collapse together at $p=p^\mu_{c1}$,
while for $q_{12}\leq 1-\frac{p_c}{p^\mu_{c1}}$, $p_{eff}>p_c$ and the second lattice undergoes a second order transition.
In the second scenario from Eq. (\ref{eq_p_eff}) we obtain that
\begin{equation}\label{p_mu_mu}
p^{\mu}_{c2}=p_c/(1-q_{12}).
\end{equation}

The general results obtained analytically for two coupled lattices are also useful for analyzing network-of-lattices.
For any given structure of the network-of-lattices, the first lattice that breaks down will undergo a first-order-transition.
The reason is that the behavior of all other lattices connected to this lattice can be reduced to a single "effective"
percolation function $g_{eff}(x)$ and the considerations of two coupled lattices are valid.
Once one or several lattices  collapse the survived lattices that become isolated will undergo a second-order transition.

\begin{figure}[h]
    \begin{center}
        \subfigure{\includegraphics[width = 0.4\textwidth]{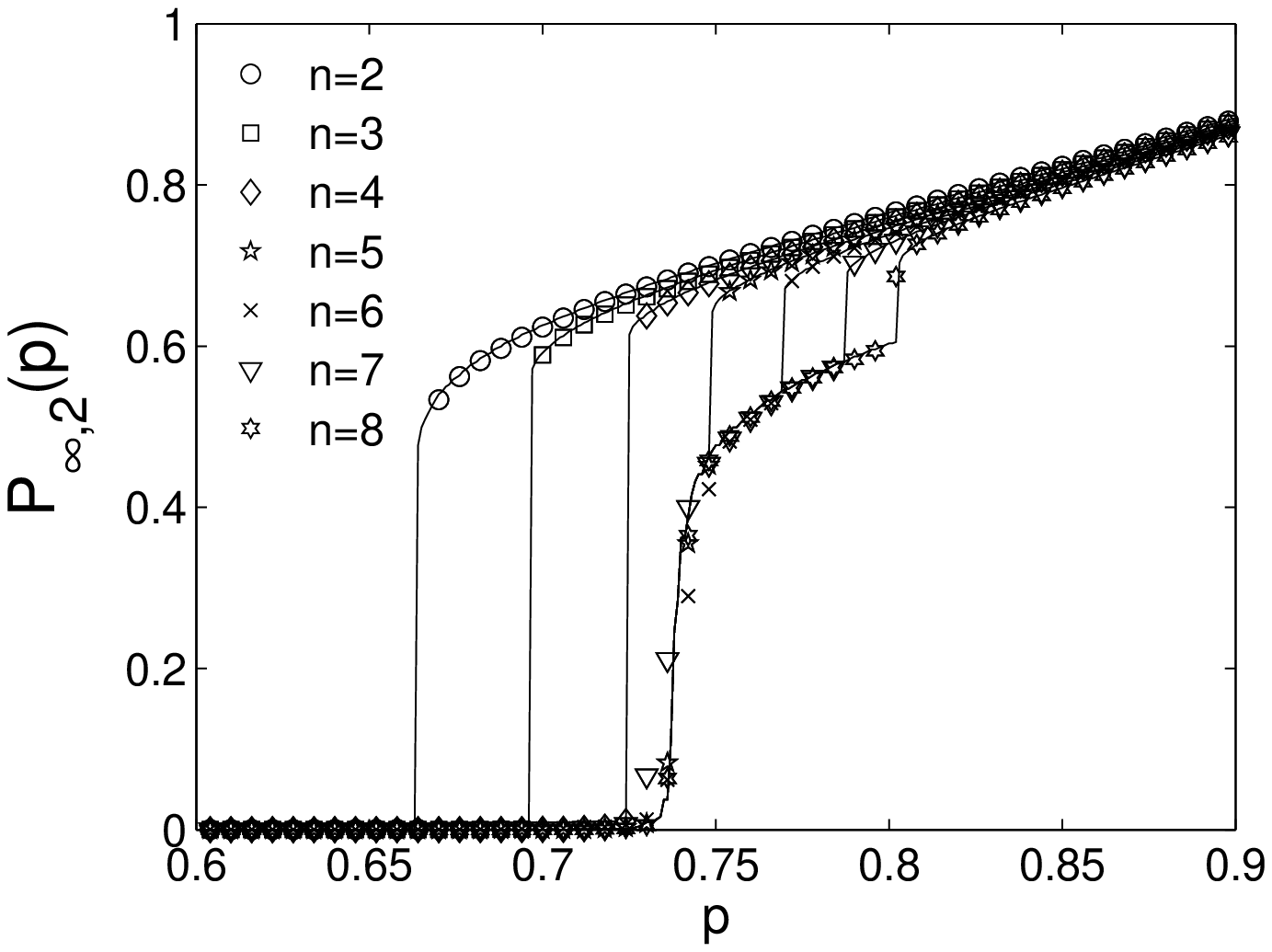}}
        \subfigure{\includegraphics[width = 0.4\textwidth]{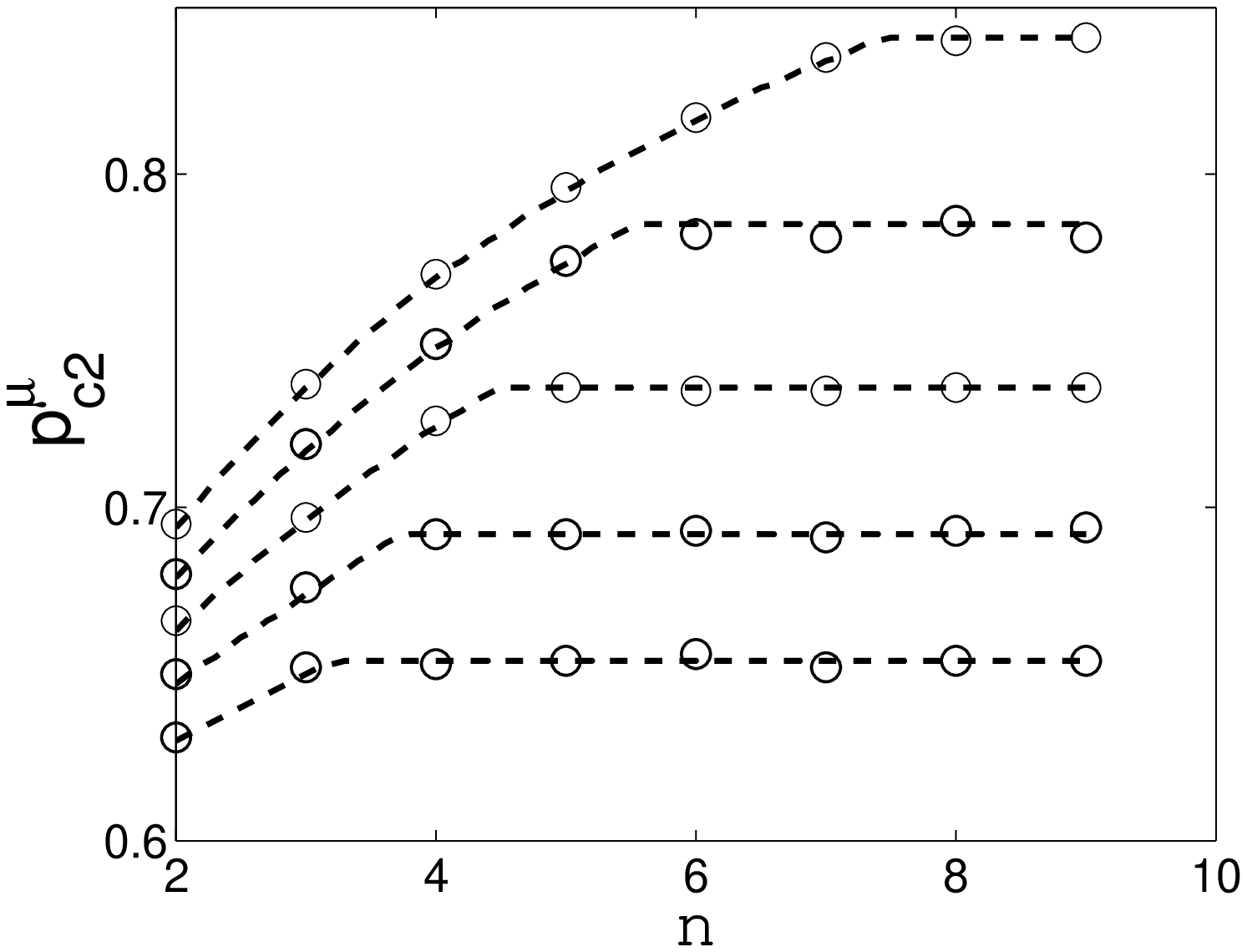}} 	
    \end{center}
    \caption{\textbf{Percolation in a starlike network-of-lattices. a,}
    The size of the peripheral lattices $P_{\infty,2}(p)$ in a starlike network of $n$ lattices with $q=0.2$ as obtained analytically from Eqs. (\ref{NetOfLattices1}) and (\ref{NetOfLattices2}) (solid lines) and  from numerical simulations (symbols).
    The discontinuous jumps in the peripheral lattices occur when the root lattice collapses at $p=p^\mu_{c1}$.
    For small $n$, the peripheral lattices collapse together with the root, $p^\mu_{c2}=p^\mu_{c1}$, while for large
    $n$, the peripheral lattices remain functional even after the root lattice collapses and continuously disintegrate at
    $p^\mu_{c2}<p^\mu_{c1}$, which is given by Eq. (\ref{p_mu_mu}).
    \textbf{b,} The critical threshold of the peripheral lattices $p^\mu_{c2}$ versus the number of lattices $n$ for different values
    of coupling strength $q$ from bottom to top: $q=0.1$, $1.5$, $2$, $2.5$, $3$. The dashed lines are the theory and the points are from simulations.
    \label{p_inf_vs_n}}
\end{figure}

Next, we analyze analytically and numerically a system of starlike network-of-lattices.
The system consists of $n$ lattices with equal dependency, $q_{ij}=q$ for $i=1$ or $j=1$ and $q_{i,j}=0$ elsewhere.
A fraction of $1-p$ of the nodes is randomly removed from all the lattices, $p_i=p$.
The steady state of the system is given by
\begin{eqnarray}
  x_1 &=& p[q p g(x_2)-q+1]^{n-1}  \label{NetOfLattices1}\\
  x_2 &=& q p^{\frac{n}{n-1}} x_1^{\frac{n-2}{n-1}}g(x_1)+p(1-q)
  \label{NetOfLattices2}
\end{eqnarray}
where the giant component of the root lattice is $P_{\infty,1}=x_1g(x_1)$ and the giant component of the $n-1$
peripheral lattices is $P_{\infty,2}=x_2g(x_2)$ \cite{Gao2011}.
In the case of $n=2$ Eqs. (\ref{NetOfLattices1}) and (\ref{NetOfLattices2}) reduce to Eqs. (\ref{eq_symmetric}) and (\ref{eq_symm_qc}). For $n>2$, the root
lattice will collapse first since it depends on all the other lattices, consequently, it undergoes first-order transition
for any value of $q$. Figure \ref{p_inf_vs_n} shows the percolation behavior of the peripheral lattices in the starlike network-of-lattices.
We find that for small $n$ the peripheral lattices collapses together with the root lattice at $p=p^\mu_{c1}$.
However, for large $n$, the collapse of the root is followed by a discontinuous fall in $P_{\infty,2}$ at $p=p^\mu_{c1}$ which then
undergoes a second-order transition at $p^{\mu}_{c2}<p^\mu_{c1}$.

\section{Non-symmetric interdependent networks}
\label{NonSymmSect}
Here we analyze the mutual percolation of any two interdependent networks for the non-symmetric case and find the conditions for discontinuous
behavior.
The steady state is given by the set of two equations \cite{Gao2011}
\begin{eqnarray}
x_1 &=& p_1p_2q_{21}g_2(x_2)+p_1(1-q_{21}) \label{eq_Gao1}\\
x_2 &=& p_2p_1q_{12}g_1(x_1)+p_2(1-q_{12}) \label{eq_Gao2}
\end{eqnarray}
where the size of the giant component of network $i$ is given by $P_{\infty,i}=x_ig_i(x_i)$,
$p_i$ is the initial random occupation of network $i$ and $q_{ij}$ is the fraction of nodes in network $j$
that depend on network $i$. For given $g_1(x)$ and $g_2(x)$, Eqs. (\ref{eq_Gao1}) and (\ref{eq_Gao2})
can be solved for $x_1$ and $x_2$, for any values of $p_i$ and $q_{ij}$.
We assume, for simplicity, that in both networks a fraction $1-p$ is initially randomly removed, $p_1=p_2=p$.
Still, since $q_{12}\neq q_{21}$ and $g_1\neq g_2$ the two networks generally will not disintegrate together.

We define network $1$ as the network that first collapses when $p=p^\mu_c$, thus, the solution of $x_1$ discontinuously jumps from
$x_1>x_{c1}$ for $p>p^\mu_c$ to $x_1<x_{c1}$ for $p<p^\mu_c$. For $p=p^\mu_c$ $x_1=x^\mu$.
$g_2(x)$ can be expanded using Tailor series around $\tilde{x}_2$, the value of $x_2$ at $p \rightarrow p^\mu_c$,
\begin{equation}\label{eq_x2_Tailor}
    g_2(x)=g_2(\tilde{x}_2)+g'_2(\tilde{x}_2)(x-\tilde{x}_2)+\ldots
\end{equation}
For $p\rightarrow p^\mu_c$, Eq. (\ref{eq_Gao1}) becomes,
$x_1 = p^2q_{21}[g_2(\tilde{x}_2)+g'_2(\tilde{x}_2)(x_2-\tilde{x}_2)]+p(1-q_{21})$,  or
\begin{equation}\label{eq_x2ofx1}
    x_2=mx_1+n
\end{equation}
where $m=\frac{1}{p^2q_{21}g'_2(\tilde{x}_2)} $ and $n=\tilde{x}_2-p(1-q_{21})-\frac{g_2(\tilde{x}_2)}{g'_2(\tilde{x}_2)} $.

Equation (\ref{eq_Gao2}) becomes
\begin{equation}\label{eq_trans_x1}
    mx_1+n=p^2q_{12}g_1(x_1)+p(1-q_{12}).
\end{equation}
At the critical point, where $p=p^\mu_c$, the solution $x_1=x_1^\mu$ satisfies both Eq. (\ref{eq_trans_x1}) and
the tangential condition, obtained by taking the derivative of Eq. (\ref{eq_trans_x1}) with respect to $x_1$
\begin{equation}\label{eg_der_x1}
    m=p^2q_{12}g'_1(x_1),
\end{equation}
which then becomes
\begin{equation}\label{eq_der_g1_g2}
    p^4q_{12}q_{21}g'_1(x_1)g'_2(\tilde{x}_2)=1.
\end{equation}

The solution of Eqs. (\ref{eq_trans_x1}) and (\ref{eq_der_g1_g2}) provides $p^\mu_c$ and $x_1^\mu$ of the
critical point of the first-order transition. Decreasing the values of $q_{12}$ and $q_{21}$ leads to lower value
for $x_1^\mu$ and, consequently, lower value for $P_{\infty,1}(p^\mu_c)=x_1^\mu g_1(x_1^\mu)$, the size of the giant
component at criticality. If $x_1^\mu \rightarrow x_{c1}$ then $P_{\infty,1}(p^\mu_c) \rightarrow 0$ and the transition becomes
second-order.

\section{Interdependent networks with feedback-dependency-links}
\label{feedback}
As mentioned earlier, the above formalism assumes "no-feedback condition", i.e., if a node $A_i$ in network $A$ depends on
a node $B_i$ in network $B$, node $B_i$ may depend only on node $A_i$ and not on any other node in network $A$.
In this section we analyse the case of interdependent networks without such a constraint. The dependency links between the networks
are chosen randomly and thus, dependency chains exist, $A_i$ depends on $B_i$ and $B_i$ depends on $A_j$ and so on. .

In this case Eq. (\ref{eq_symmetric}) becomes \cite{Gao2011}
\begin{equation}\label{eq_symmetric_feedback}
    x-p(1-q)=pqP_\infty(x).
\end{equation}
where $P_\infty(x)$ is the relative size of the giant component compared to the initial size of the network after random removal
of a fraction $1-x$ of the nodes. Equation (\ref{eq_der_symmetric}), the condition for first-order transition becomes
 \begin{equation}\label{eq_der_symmetric_feedback}
   1=pqP'_\infty(x).
 \end{equation}

\begin{figure}[h]
    \begin{center}
        \subfigure{\includegraphics[width = 0.4\textwidth]{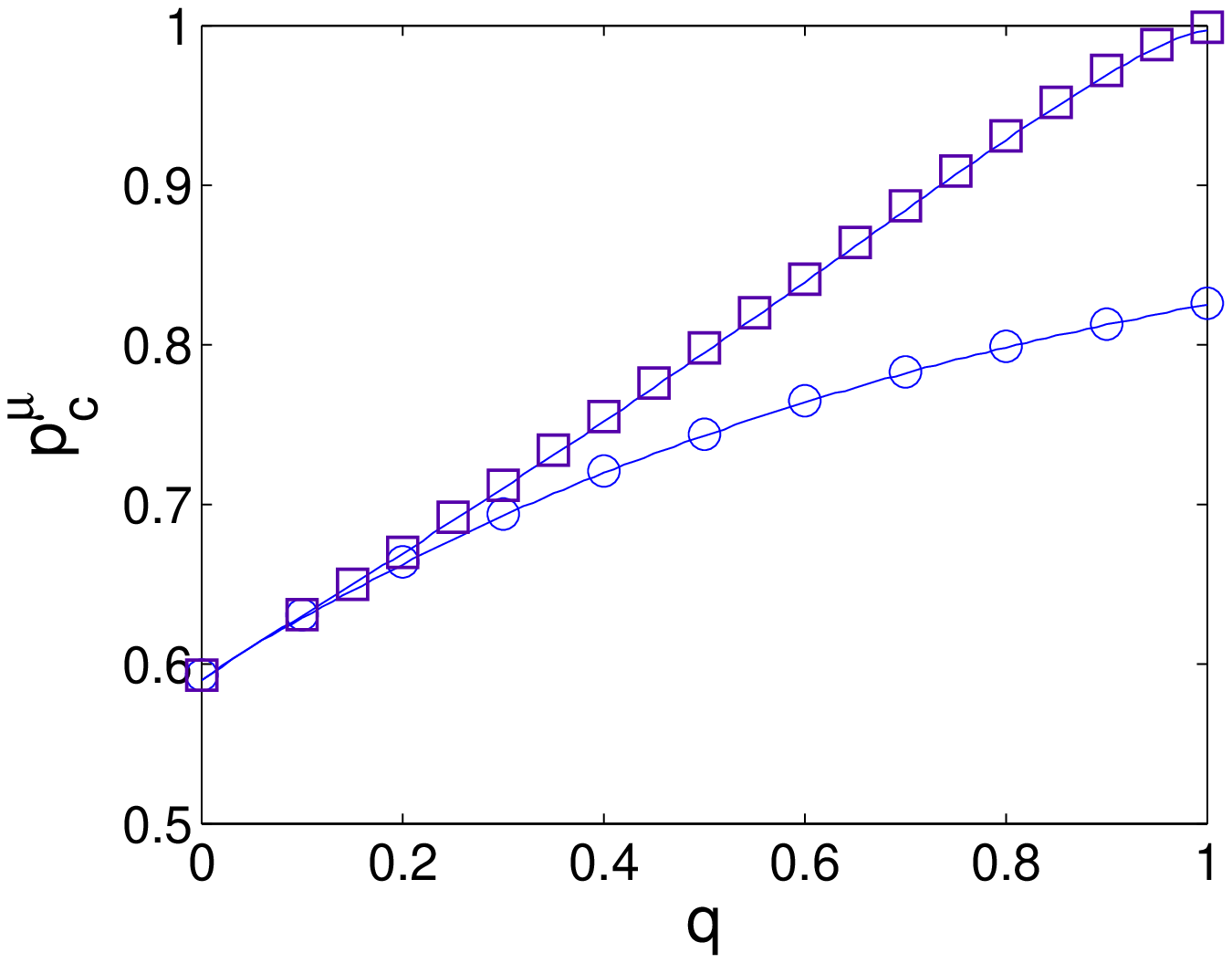}}
        \subfigure{\includegraphics[width = 0.4\textwidth]{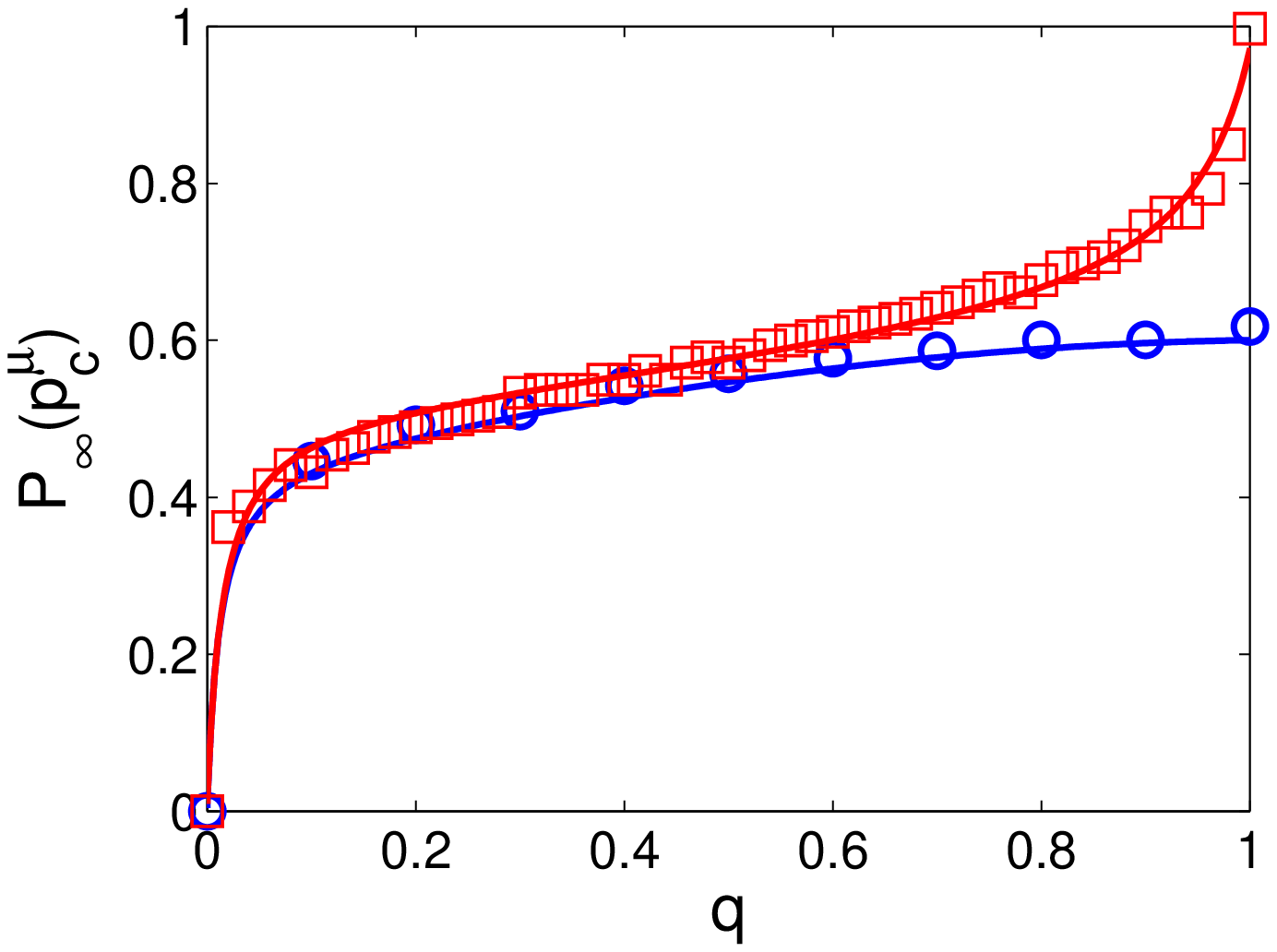}} 	
    \end{center}
    \caption{\textbf{The effect of feedback dependency links.
    a,} The critical threshold $p_c^\mu$ for the case of interdependent lattices with feedback-dependency-links (squares)
     and without feedback-dependency-links (circles). The symbols represent results of simulations and the solid lines
     are the solutions of Eqs. (\ref{eq_symmetric}) and (\ref{eq_der_symmetric}) for the case of no-feedback condition,
     and of Eqs. (\ref{eq_symmetric_feedback}) and (\ref{eq_der_symmetric_feedback}) for the case of feedback condition.
     \textbf{b,} The size of the collapse in interdependent lattices with feedback-dependency-links (squares) compared to the case
     of no-feedback condition (circles).
    \label{FeedbackVsNofeedback}}
\end{figure}

Figure \ref{FeedbackVsNofeedback} compares the case of interdependent lattices with feedback-dependency-links to the case
of no-feedback condition, which is discussed above in section \ref{SymmNofeedback}. It is seen that the feedback case is more vulnerable compared to the no feedback case. The critical threshold and the size of the giant component at criticality are both larger compared to the no-feedback rule. This can be understood as follows. For small values of $q$, the probability
to find a dependency chain of length $l$ exponentially decreases, $p(l)\sim q^l$ and the dependency chains have a characteristic length,
 thus, the results are similar to the case with no-feedback condition.
 The length of the dependency chains diverges for $q\rightarrow 1$, where dependency chains of the size of the system appear and the system become
 extremely unstable, $p_c^\mu=1$.

Similarly, the solution of a starlike network-of-networks, given before by Eqs. (\ref{NetOfLattices1}) and (\ref{NetOfLattices2}),
 becomes for the case of feedback dependency

\begin{eqnarray}
x_1 &=& p[qP_\infty(x_2)-q+1]^{n-1}  \label{NetOfLattices1_feedback}\\
  x_2 &=& pqP_\infty(x_1)+p(1-q)
  \label{NetOfLattices2_feedback}
\end{eqnarray}
where the giant component of the root lattice is $P_\infty(x_1)$ and the sizes of the giant components of the $n-1$
peripheral lattices is $P_\infty(x_2)$.

\FloatBarrier


\end{document}